\documentclass[dvips]{acta}
\usepackage{supertabular,lscape,epsfig}
\usepackage{amssymb}
\usepackage{amsmath}

\newcommand{\apj}{ApJ}

\newcommand{\apjl}{ApJL}
\newcommand{\aj}{AJ}
\newcommand{\aap}{A\&A}

\newcommand{\mnras}{MNRAS}
\newcommand{\pasp}{PASP}

\SetPages{0}{0}

\SetVol{69}{2019}

\begin{document}

\begin{Titlepage}
\Title{CoRoT-18 b: Analysis of high-precision transit light curves with starspot features}
\Author{St. Raetz$^{1,2,3}$, A. M. Heras$^{3}$, P. Gondoin$^{4}$, M. Fern\'{a}ndez$^{5}$, V. Casanova$^{5}$, T.O.B. Schmidt$^{6}$ and G. Maciejewski$^{7}$}
{$^{1}$Institute for Astronomy and Astrophysics T\"{u}bingen (IAAT), University of T\"{u}bingen, Sand 1, D-72076 T\"{u}bingen, Germany \\
e-mail:raetz@astro.uni-tuebingen.de\\
$^{2}$Freiburg Institute of Advanced Studies (FRIAS), University of Freiburg, Albertstra\ss{}e 19, D-79104 Freiburg, Germany\\
$^{3}$Science Support Office, Directorate of Science, European Space Research and Technology Centre (ESA/ESTEC), Keplerlaan 1, 2201 AZ Noordwijk, The Netherlands\\
$^{4}$Future Missions Department, Directorate of Science, European Space Research and Technology Centre (ESA/ESTEC), Keplerlaan 1, 2201 AZ Noordwijk, The Netherlands\\
$^{5}$Instituto de Astrof\'{\i}sica de Andaluc\'{\i}a, CSIC, Apdo. 3004, 18080 Granada, Spain\\
$^{6}$Hamburger Sternwarte, Gojenbergsweg 112, D-21029 Hamburg, Germany\\
$^{7}$Centre for Astronomy, Faculty of Physics, Astronomy and Informatics, Nicolaus Copernicus University, Grudziadzka 5, 87-100 Torun, Poland 
}

\Received{April 12, 2019}
\end{Titlepage}

\Abstract{When a planet occults a spotty area on a stellar surface, the flux increases and a characteristic feature in a light curve - a bump - is observed. Among the planets detected by the \textit{CoRoT}-mission CoRoT-18 is especially interesting as it exhibited spot crossings that we have analysed in detail. We used four ground-based observations obtained at a 1.5-m telescope in Spain and the 13 available \textit{CoRoT}-transits to refine and constrain stellar, planetary and geometrical parameters of the system. We found that the derived physical properties slightly deviate from the previously published values, most likely due to the different treatment of the stellar activity.\\ Following a spot over several transits enabled us to measure the stellar rotation period and the spin-orbit alignment. Our derived values of $P_{\mathrm{rot}}=5.19\pm0.03$\,d and $\lambda=6\pm13^{\circ}$ are in agreement with the literature values that were obtained with other methods. Although we cannot exclude a very old age for CoRoT-18, our observations support the young star hypothesis and, hence, yield constraints on the time-scale of planet formation and migration.}{planets and satellites: individual: CoRoT-18\,b -- stars: individual: CoRoT-18  -- planetary systems.}

\section{Introduction}

The \textit{CoRoT} satellite, which was launched in 2006 and decommissioned in 2013, provided six years of high precision photometry of a very high number of stars. So far 37 \textit{CoRoT}  confirmed exoplanets have been published while further 557 transiting planet candidates are being screened for confirmation (Deleuil et al. 2018). The majority of the discovered planets have the size of Jupiter, except CoRoT-8b, which appears to be a hot Neptune (Bord\'{e} et al. 2010), and CoRoT-7b which is the first rocky Super-Earth ever detected (L\'{e}ger et al. 2009).\\Because of its orbit around the Earth, the \textit{CoRoT} satellite could continuously observe one selected target field for up to 150\,days. Therefore ground-based follow-up observations are needed to refine the orbital elements of the planets, constrain the physical parameters of the planets and stars and to search for additional bodies in the system.\\ Three stars of the \textit{CoRoT} sample (CoRoT-2, 18 and 20) are of special importance since they appear to be younger than 1\,Gyr  (Guillot \& Havel 2016). Since most transit surveys are biased towards non-active stars to favour transit detection and radial-velocity follow-up measurements, almost all planet host stars and hence also their planets are Gyrs old. However, the observation and analysis of young systems provide important constraints on the time-scale of planet formation and migration.\\ While there are numerous studies on CoRoT-2 (e.g. Wolter et al. 2009; Silva-Valio et al. 2010; Silva-Valio \& Lanza 2011; Bruno et al. 2016), CoRoT-18 has not received enough attention. This is surprising since these two systems are almost twins. Both stars show a complex global light curve (LC) which hints at the hypothesis that they are heavily spotted. Furthermore, both systems are particularly interesting because of their fast stellar spin and the presence of a close-in, relatively massive giant planet. To enhance our understanding of these so far rarely detected young Hot Jupiters, we launched a dedicated follow-up campaign for CoRoT-18. Moreover, high-precision transit light curves of a planet moving across a spotted stellar disk offer an outstanding opportunity to map the starspot distribution  (Schneider 2000; Silva 2003) and, hence, allows us to probe the stellar surface.

\begin{table}
\centering
\caption{Physical and orbital properties of the CoRoT-18\,b system summarised from literature.}
\begin{tabular}{lr@{\,$\pm$\,}l|r@{\,$\pm$\,}l}
\hline \hline
Parameter & \multicolumn{2}{c|}{Value} & \multicolumn{2}{c}{Raetz et al. (2019)} \\ \hline
Epoch zero transit time $T_{0}$  [d] & \multicolumn{2}{l|}{2455321.72412 [1]} & \multicolumn{2}{l}{2455321.72565} \\
 & & 0.00018 [1] & & 0.00024 \\
Orbital period $P$  [d] & \multicolumn{2}{l|}{1.9000693 [1]} & \multicolumn{2}{l}{1.9000900}  \\
 & & 0.0000028 [1] & & 0.0000005 \\
Semi-major axis $a$ [au] & 0.02860 & 0.00065 [2] & 0.0288 & 0.0008\\
Inclination $i$  [$^{\circ}$] & 86.5 & $^{1.4}_{0.9}$ [1] & 89.9 & $^{1.6}_{1.6}$ \\
Eccentricity $e$ & 0.10 & 0.04 [3] & \multicolumn{2}{c}{} \\
Impact parameter $b$ & 0.40 & $^{0.08}_{0.14}$ [1] & 0.01 & $^{0.20}_{0.20}$ \\
Mass star $M_{\mathrm{A}}$  [M$_{\odot}$] & 0.861 & 0.059 [2] & 0.88 & 0.07 \\
Radius star $R_{\mathrm{A}}$  [R$_{\odot}$] & 0.924 & 0.057 [2] & 0.883 & $^{0.025}_{0.031}$ \\
Density star $\rho_{\mathrm{A}}$  [$\mathrm{\rho}_{\mathrm{\odot}}$]& 0.96 & 0.17 [1] & 1.28 & $^{0.04}_{0.09}$\\
Eff. temperature $T_{\mathrm{eff}}$  [K] & 5440 & 100 [1] & \multicolumn{2}{c}{} \\
Surface gravity star log$\,g_{\mathrm{A}}$ & 4.442 & 0.043 [2] & 4.491 & $^{0.015}_{0.023}$  \\
Metallicity $\left[ \frac{Fe}{H}\right] $ & -0.1 & 0.1 [1] & \multicolumn{2}{c}{}  \\
Stellar luminosity log$\frac{L_{\mathrm{A}}}{L_{\mathrm{\odot}}}$ & \multicolumn{2}{c|}{} & -0.17 & 0.06 \\
Stellar age log(Age) & \multicolumn{2}{c|}{} & 7.50 & 0.04\\
&  \multicolumn{2}{c|}{} & 9.84 & 0.26 \\
Stellar rotation period $P_{\mathrm{rot}}$ [d] & 5.4 & 0.4 [1] & \multicolumn{2}{c}{} \\
Mass planet $M_{\mathrm{b}}$  [$M_{\mathrm{Jup}}$] & 3.27 & 0.17 [2] & 3.30 & $^{0.19}_{0.19}$ \\
Radius planet $R_{\mathrm{b}}$  [$R_{\mathrm{Jup}}$] & 1.251 & 0.083 [2] & 1.146 & $^{0.039}_{0.048}$ \\
Density planet $\rho_{\mathrm{b}}$  [$\mathrm{\rho}_{\mathrm{Jup}}$]  & 1.56 & 0.30 [2] & 2.06 & $^{0.24}_{0.29}$\\
Surface gravity planet log\,$g_{\mathrm{b}}$ & 3.714 & 0.055 [2] & 3.797 & $^{0.021}_{0.030}$ \\
Eq. temperature $T_{\mathrm{eq}}$ [K] & 1490 & 45 [2] & 1487 & $^{19}_{19}$ \\
Safronov number $\Theta$ & 0.173 & 0.012 [2] & 0.189 & $^{0.019}_{0.020}$ \\
Distance $d$ [pc] & 870 & 90 [1] & \multicolumn{2}{c}{}\\ 
Spectral type &  \multicolumn{2}{c|}{G9V [1]}  & \multicolumn{2}{c}{}\\
\hline \hline
\end{tabular}
\\ References: [1] H\'{e}brard et al. (2011), [2] Southworth (2012), [3] Parviainen et al. (2013)
\end{table}

\section{CoRoT-18}

CoRoT-18\,b, that orbits its G9V host star in $\sim$1.9\,d, was detected in the third short run of \textit{CoRoT} in the Galactic anti-centre direction (SRa03, H\'{e}brard et al. 2011). With a mass of $\sim$3.4\,$M_{\mathrm{Jup}}$, a radius of $\sim$1.3\,$R_{\mathrm{Jup}}$ and a slight non-zero (e\,$<$\,0.08) eccentricity, it belongs to the group of massive Hot Jupiter planets on elliptical orbits. Parviainen et al. (2013) measured an eccentricity of e\,=\,0.10\,$\pm$\,0.04 by observing a marginally significant secondary eclipse event. CoRoT-18 appears to be a young star when considering the stellar activity, the lithium abundance and the stellar spin. However, the values obtained from evolution tracks do not exclude very old ages. The follow-up LC of CoRoT-18 observed by H\'{e}brard et al. (2011) shows a starspot feature during the transit and, hence, supports the hypothesis of a young stellar age. The presence of in-transit starspot features was confirmed by Raetz et al. (2019). By plotting CoRoT-18 in a modified Hertzsprung-Russel diagram together with isochrones, Raetz et al. (2019) confirmed that CoRoT-18 is consistent with very young and old ages.\\ In Raetz et al. (2019) all starspot features in the transit were removed from the LCs before the transit modelling. The derived physical properties were found to  deviate slightly from the previously published values, most likely due to the different treatment of the stellar activity. Raetz et al. (2019) improved the transit ephemeris and found an orbital period that is 1.8\,s longer and six times more precise than the previous published value. They could not find evidence for transit timing variations.\\ Table~1 summarises the system parameters known from the literature and the properties derived by Raetz et al. (2019) which include the equilibrium temperature of the planet $T_{\mathrm{eq}}$ (assuming a Bond albedo\,=\,0 and only little energy redistribution across the surface of the planet; Hansen \& Barman 2007) and the Safronov number $\Theta$. The Safronov number is defined as the square of the ratio of escape velocity of the planet and orbital velocity which is a measure of the efficiency with which a planet gravitationally scatters other bodies (Safronov 1972).

\section{Observation, data reduction and photometry}

Our dedicated ground-based follow-up campaign of CoRoT-18 started in 2014. Since our first observation in 2014 January showed evidence of the passage of the planet over a starspot, we continued collecting high-precision transit LCs of CoRoT-18. Our ground-based observations range from epoch 714 to 1104, where epoch $E$ is related to the linear transit ephemeris by the expression $T_{\mathrm{Transit}}=T_{0}+P\cdot E$ where P and $T_{0}$ are respectively the period and the mid-transit time to epoch zero listed in Table~1. Moreover we re-analysed the data taken by the \textit{CoRoT} mission and paid special attention to any signs of stellar spots during the transits. We also re-analysed publicly available stellar spectra to investigate stellar properties and study the activity. The detailed information of the observations and analysis is given in the following sections. The final LCs are shown in Fig.~1 and 2.

\begin{figure}
 \centering
  \includegraphics[width=0.9\textwidth]{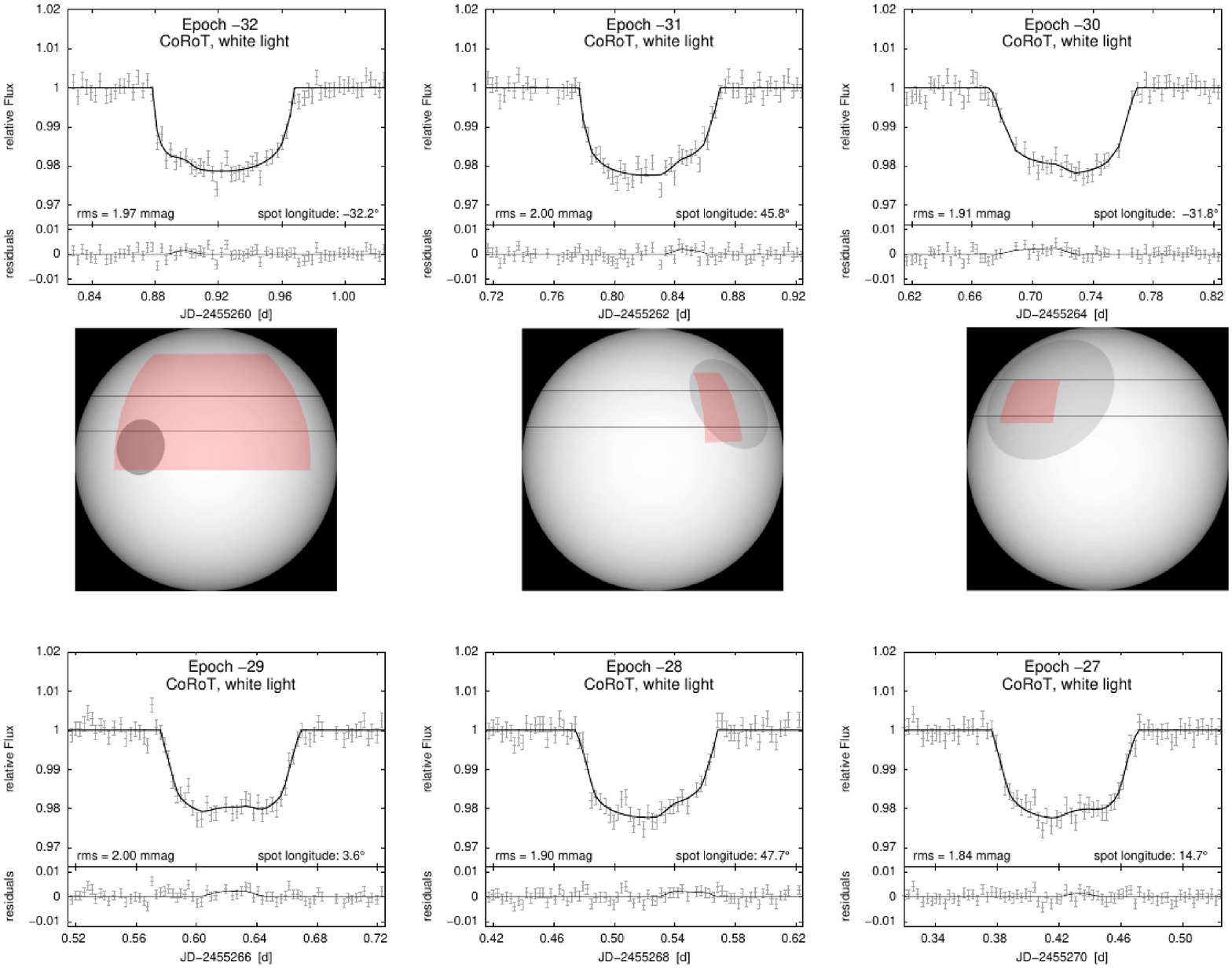}
  \includegraphics[width=0.9\textwidth]{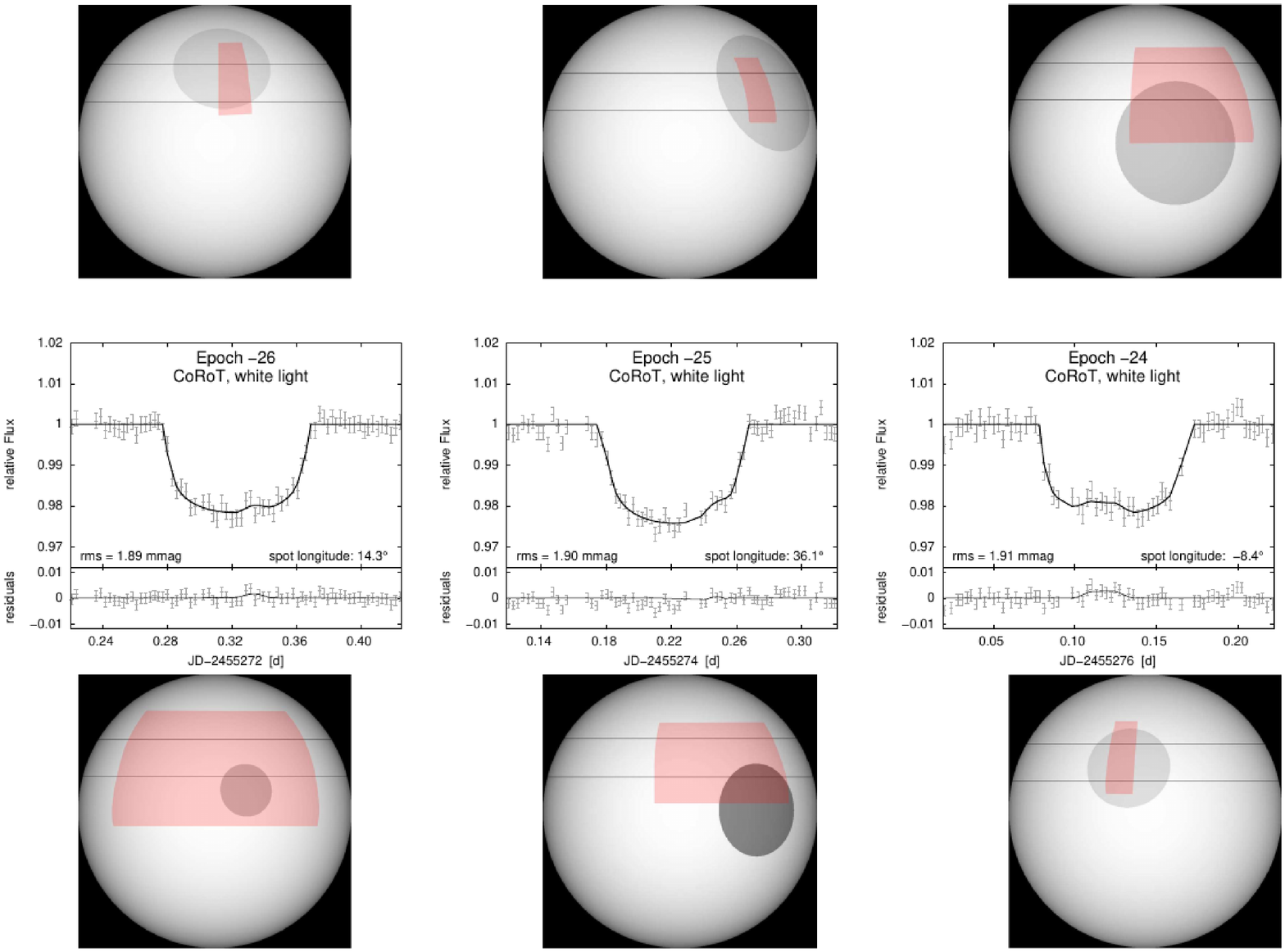}
  \caption{LCs of CoRoT-18\,b with best-fitting spot model and their residuals. The theoretical line in the residual panel gives the difference between the spot- and the spot-free-model. Also shown is a representation of the stellar disc with the spot that corresponds to the best-fitting model. Note that this is just one instance of the spot models used in the derivation of the spot parameters. The transit path is represented by the two horizontal black lines. The red shaded area show the range of spot centre locations of the final solution (Table~3). }
\label{LC_CoRoT18a}
\end{figure}

\begin{figure}
 \centering
  \includegraphics[width=0.9\textwidth]{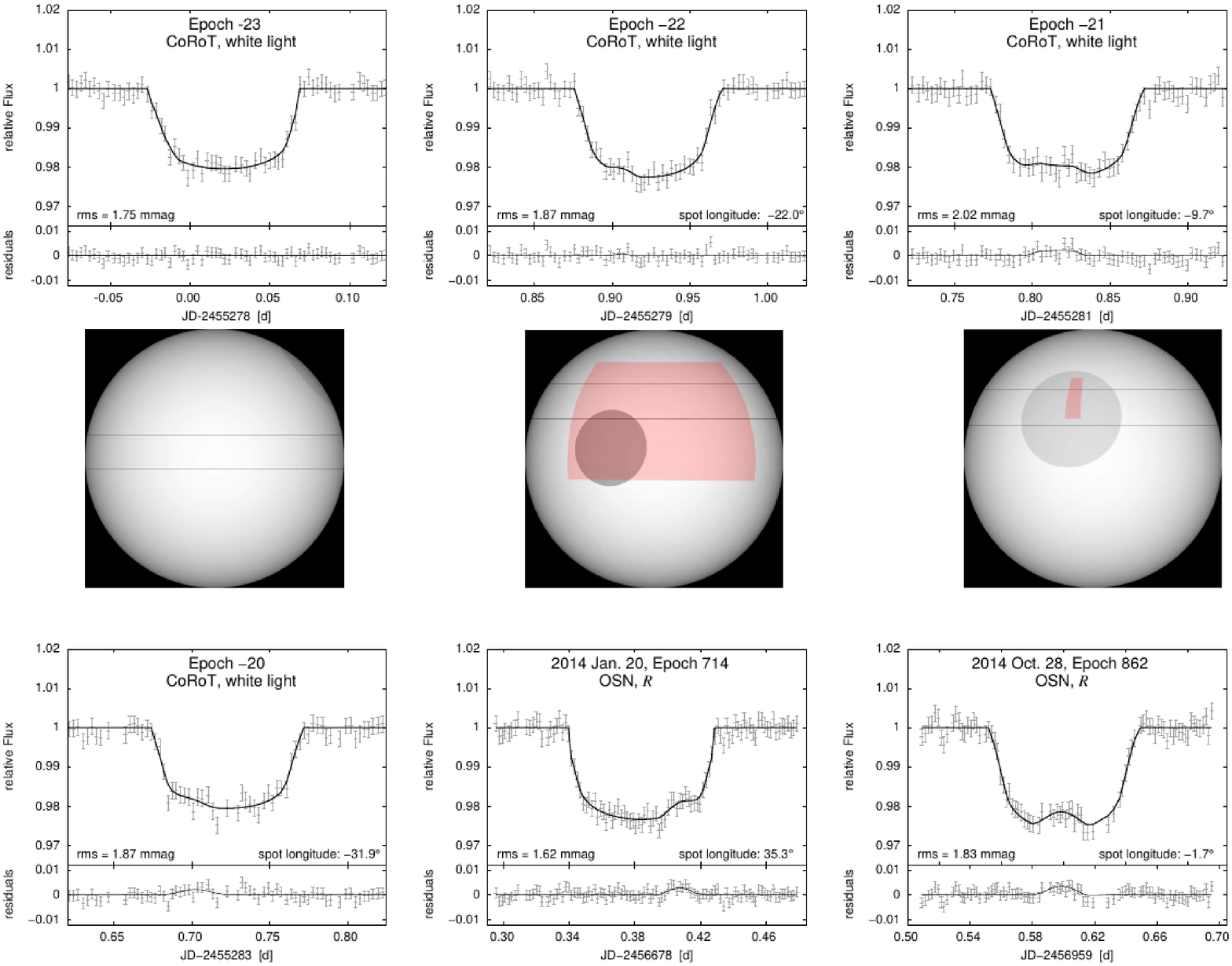}
  \includegraphics[width=0.9\textwidth]{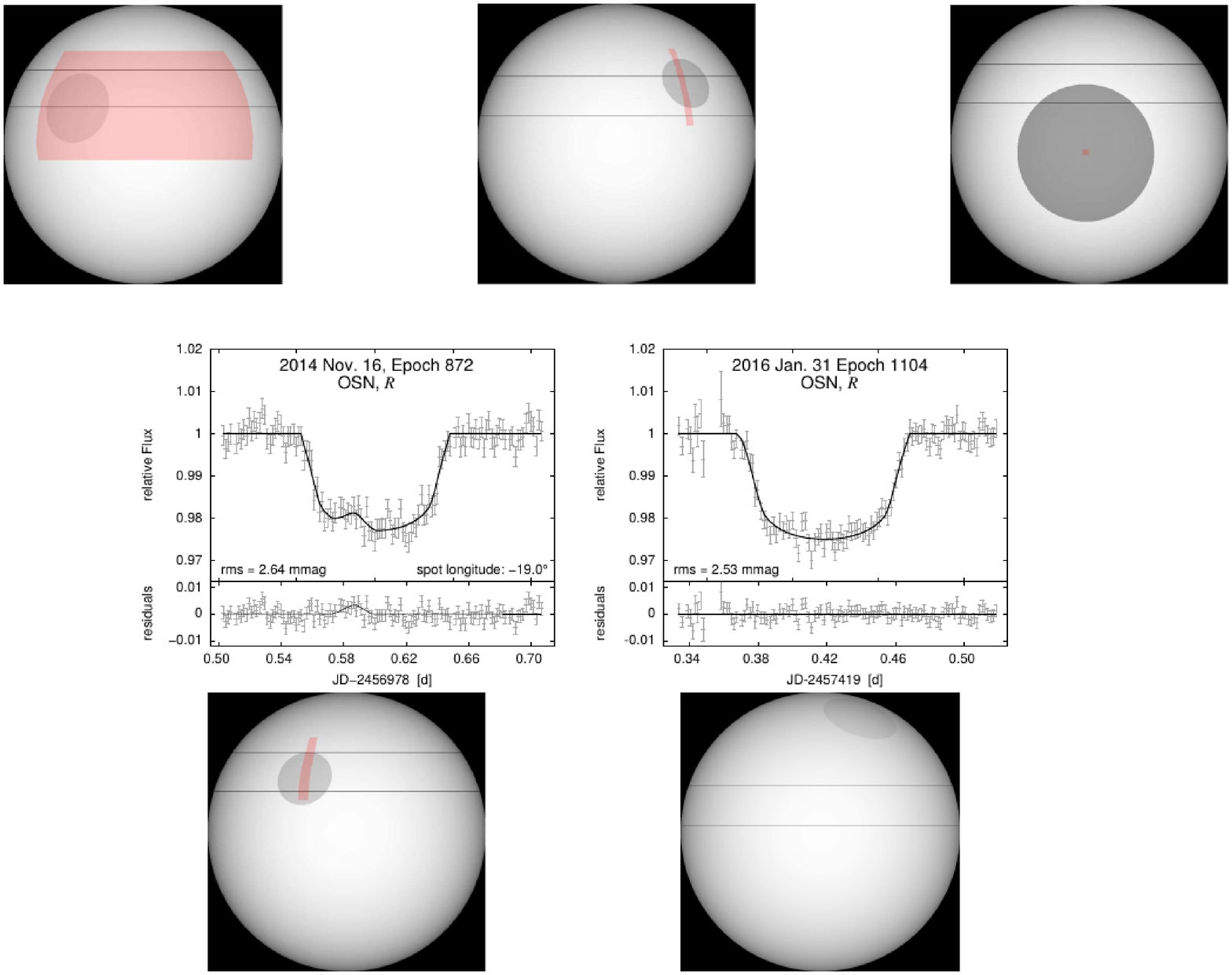}
  \caption{Same as Fig.~1.  For the \textit{CoRoT} LC at epoch -23 and OSN LC from 2016 Jan. 31st (epoch 1104) the PRISM best-fitting model only shows a spot feature outside of the transit path. Since these spot locations are random due to the lack of in-transit information and the high level of degeneracy we do not give any errors.}
\label{LC_CoRoT18b}
\end{figure}

\subsection{\textit{CoRoT} observations}

CoRoT-18 was observed by \textit{CoRoT} from 2010 March 5 to 29. The cadence was 32\,s throughout the observations. The data were downloaded from the ``NASA Exoplanet Archive'' (Akeson et al. 2013, http://exoplanetarchive.ipac.caltech.edu/) which provides fully reduced photometric time-series. The available public data from \textit{CoRoT} are the N2 (the primary scientific, Version 2.2) data as produced by the \textit{CoRoT} pipeline (Auvergne et al. 2009). The original white-light LC which was used for the analysis consists of 65120 exposures. After removing all flagged measurements (flagged e.g. because of energetic particles, South Atlantic Anomaly crossings, Earth eclipses; Chaintreuil et al. 2016) we were left with 56823 data points including 13 transit events. The time stamp was corrected from the heliocentric julian date at the end of the measurements to the middle of the exposure. We extracted the transit events by using all data points $\pm$0.2\,d around the expected transit time. The transit LCs were normalised by dividing by the average out-of-transit flux. \\ The observed star fields are sometimes quite crowded and, hence, a number of objects are likely to fall within the large aperture around CoRoT-18 resulting in additional light (``third'' light, $L_{\mathrm{3}}$) that contaminates the photometry. H\'{e}brard et al. (2011) found a contamination factor of $L_{\mathrm{3}}\,=\,2.0\pm0.1\%$ for CoRoT-18. We subtracted $L_{\mathrm{3}}$ from the normalised flux before re-normalising. As a preparation for the analysis we applied $\sigma$-clipping and we removed photometric trends by fitting a second-order polynomial to the out-of-transit LC.\\ Finally, the transit LCs were binned by a factor of seven (equivalent to 224\,s cadence). This shortened the computing time for spot modelling and reduced the noise. Enough in-transit data points were kept to avoid losing information. 


\subsection{Ground-based observations}

We observed four transit events in 2014 and 2016 with a 1.5-m reflector at the Observatorio de Sierra Nevada (OSN) which is operated by the Instituto de Astrof\'{i}sica de Andaluc\'{i}a, CSIC, Spain. Detailed information on the used instrument and the observations are given in Raetz et al. (2019). All observations were performed in $R$-band.\\ We carried out the data reduction and photometry in the same way as in Raetz et al. (2015, 2016). For the basic data reduction (subtraction of bias and division by a sky flat field) we used the \begin{scriptsize}IRAF\end{scriptsize}\footnote{\begin{scriptsize}IRAF\end{scriptsize} is distributed by the National Optical Astronomy Observatories, which are operated by the Association of Universities for Research in Astronomy, Inc., under cooperative agreement with the National Science Foundation.} routines \textit{zerocombine}, \textit{flatcombine} and \textit{ccdproc}. To create the LCs we performed aperture photometry with \begin{scriptsize}IRAF\end{scriptsize} using ten different aperture radii. An optimised artificial comparison star  (Broeg et al. 2005) was used to produce the final differential magnitudes. The optimal aperture radius was determined by comparing the LC standard deviations for a sample of constant stars.\\ As also done for the \textit{CoRoT} LCs we initially applied $\sigma$-clipping and corrected photometric trends. As a last step we converted the magnitudes into fluxes and normalised the LC using the average value of the out-of-transit data points.

\section{Spectral analysis of CoRoT-18}

For the spectral analysis of CoroT-18 we retrieved four HARPS spectra from the ESO archive\footnote{Based on observations collected at the European Southern Observatory under ESO programme 184(C)-0639(A).} processed with the standard pipeline. The observations were obtained on 2010 December 08, 09, 17, and 18, all with an exposure time of 3600\,s and a resolution ($\lambda/\Delta\lambda$) of 115000. \\Firstly, we determined the stellar parameters. For this purpose we used the library of optical spectra of 404 touchstone stars observed with Keck/HIRES by the California Planet Search, and the tool \begin{scriptsize}Empirical SpecMatch\end{scriptsize}, which parametrises unknown spectra by comparing them against the spectral library (Yee et al. 2017). The high resolution (R$\sim$60,000) of the spectra in this library makes them more adequate for the comparison with the HARPS spectra, and therefore enables a more precise determination of the stellar parameters. The results of the \begin{scriptsize}Empirical SpecMatch\end{scriptsize} tool for the HARPS spectrum taken of 2010 December 18 are given in Table~2. These results are in agreement with the parameters derived by H\'{e}brard et al. (2011).\\ In addition, we analysed the activity of the star by examining the Ca II H and K lines. The four HARPS spectra were combined to obtain a high signal-to-noise ratio (S/N). Fig.~3 shows the resulting combined spectrum in comparison with the solar spectrum on this region (BASS2000, Observatoire de Paris, e.g. Meunier et al. 2006, http://bass2000.obspm.fr/solar\_spect.php). As can be seen, the CoRoT-18 flux in the Ca II H and K line spectral regions is higher than in the Sun, indicating a higher level of chromospheric activity. After subtracting the solar spectrum from the CoRoT-18 spectrum, the emission of the K and H lines is clearly displayed.\\ To quantify the stellar activity we calculated the $R'_{\mathrm{HK}}$ index, which is defined as the ratio of the emission from the chromosphere in the cores of the Ca II  H and K lines to the total bolometric emission of the star (Noyes et al. 1984).  For the calculation of the $R'_{\mathrm{HK}}$ index, we determined the S-index from the HARPS combined spectra, following the method described in Lovis et al. (2011). The S-index is defined as being proportional to ($H$ + $K$)/($R$ + $V$),  where $H$, $K$, $R$, and $V$ represent the total fluxes in the respective passbands. Taking a value of ($B\,-\,V$)\,=\,0.8, we obtained log $S_{\mathrm{HARPS}}$\,=\,-0.391, log $S_{\mathrm{MW}}$\,=\,-0.331 and log $<R'_{\mathrm{HK}}>$\,=\,-4.398, where $S_{\mathrm{MW}}$ represents the Mount Wilson S-index. These values are consistent with CoRoT-18 having a higher activity than the Sun, for which log $<R'_{\mathrm{HK}}>$\,=\,-4.96 (Hall et al. 2009). Moreover, the S-index value of CoRoT-18 indicates that it is a young star (according to Fig. 1 of Noyes et al. 1984). Given that the rotation period of CoRoT-18 is $\sim$5.4 days, the derived value of $<R'_{\mathrm{HK}}>$ is fully consistent with the values obtained for other stars with the same rotation periods (Fig. 3 of Noyes et al. 1984; Salabert et al. 2016, their Fig. 6). The calculated $R'_{\mathrm{HK}}$ index of CoRoT-18 is also consistent with an age of less than 1\,Gyr according to the magnetic activity evolution scenario of Sun-like stars derived from rotation period measurements in open clusters (see Fig. 3 in Gondoin 2018). 

\begin{table}
\centering
\caption{Stellar properties derived from the re-analysis of the HARPS spectra.}
\begin{tabular}{cr@{\,$\pm$\,}l}
\hline \hline
Parameter & \multicolumn{2}{c}{Value} \\ \hline
Effective temperature $T_{\mathrm{eff}}$  [K] & 5569 & 70  \\
Radius star $R_{\mathrm{A}}$  [R$_{\odot}$] & 0.97 & 0.10 \\
Mass star $M_{\mathrm{A}}$  [M$_{\odot}$] & 0.94 & 0.08 \\
Metallicity $\left[ \frac{Fe}{H}\right] $ & -0.08 & 0.09 \\
Surface gravity star log$\,g_{\mathrm{A}}$ & 4.47 & 0.12 \\
\hline \hline
\end{tabular}
\end{table}

\begin{figure}
  \centering
  \includegraphics[width=0.6\textwidth]{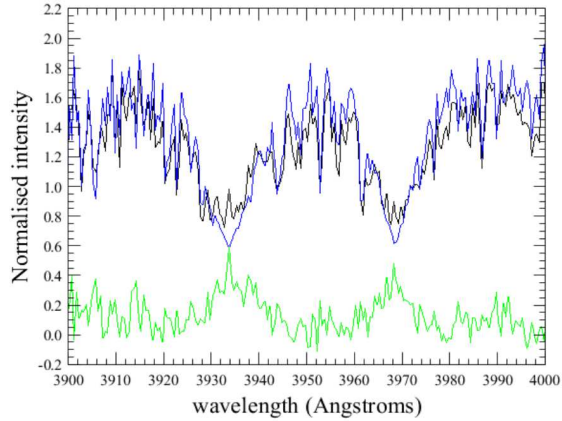}
  \caption{Combined HARPS spectrum of CoRoT-18 (black) in comparison with the solar spectrum (blue) and residuals (green) in the wavelength range around the Ca II H and K lines. The emission of the K and H lines can be clearly identified in the residuals.}
\end{figure}

\section{Rotation-Activity Relation}

A $\sim$5-day rotation period divided by a convective turnover time 18\,d $\leq\tau_{\mathrm{c}}\leq$ 24\,d for a 0.94\,$\pm$\,0.08 M$_{\odot}$ main sequence star leads to a Rossby number 0.30 $\leq Ro\leq$ 0.39. Applying the empirical relation between the $R'_{\mathrm{HK}}$ index and the Rossby number established by Mamajek \& Hillenbrand (2008), one finds -4.32 $\leq \mathrm{log}\,R'_{\mathrm{HK}}\leq$ -4.38 in good agreement with the measured emission flux $\mathrm{log}\,R'_{\mathrm{HK}}$\,=\,-4.398 in the core of the Ca II lines of CoRoT-18. \\ In view of its Rossby number, CoRoT-18 is also expected to emit X-rays in the non-saturation regime of coronal emission at an X-ray to bolometric luminosity ratio $L_{\mathrm{X}}/L_{\mathrm{bol}} \approx$ 3.6 to 7.3$\times$10$^{-5}$  (Wright et al. 2011), i.e. two order of magnitude higher than the solar value. With an effective temperature T$_{\mathrm{eff}}$\,=\,5569\,$\pm$\,70 K, this corresponds to an X-ray surface flux $F_{\mathrm{X}} \approx$ 1.8 to 4.2 $\times$ 10$^{3}$ Wm$^{-2}$. According to the parametrisation of Wood et al. (2014), such a star would emit a wind with a mass loss rate 200 to 600 times higher than the solar wind.

\section{Light curve analysis}

\subsection{Spot modelling}

The ground-based as well as the \textit{CoRoT} LCs of CoRoT-18 show features of stellar spots. The amplitudes of these signals are, on average, $\sim$4.5 times lower in the \textit{CoRoT} LCs than in the ground-based ones. In order to avoid systematic effects in the determination of the system parameters, we have simultaneously modelled the transit event and the starspots. We used the \begin{scriptsize}PRISM\end{scriptsize} (Planetary Retrospective Integrated Star-spot Model) and \begin{scriptsize}GEMC\end{scriptsize} (Genetic Evolution Markov Chain) codes developed by Tregloan-Reed et al. (2013). \begin{scriptsize}PRISM\end{scriptsize} creates a two-dimensional ``image'' of the modelled star using a pixelation approach. The anomalies in the transit LCs are modelled as single circular starspots. \begin{scriptsize}GEMC\end{scriptsize} is an optimisation algorithm, that is based on a Differential Evolution Markov Chain method and combines a genetic algorithm for global optimisation with Bayesian statistics.\\ For the LCs modelling, \begin{scriptsize}PRISM\end{scriptsize} fits geometrical parameters of the transiting system (radius ratio $k=\frac{R_{\rm{b}}}{R_{\rm{A}}}$, sum of the radii of star and planet expressed in relation to the semi-major axis $\frac{R_{\mathrm{A}}}{a}+\frac{R_{\mathrm{b}}}{a}$, orbital inclination $i$, the mid-transit time  $T_{\mathrm{c}}$, coefficients for the quadratic limb darkening law) as well as starspot parameters (longitude of the centre of the spot $\theta$, latitude of the centre of the spot $\phi$, spot size $r_{\mathrm{spot}}$, spot contrast $C_{\mathrm{spot}}$). \\ For each LC we used \begin{scriptsize}PRISM\end{scriptsize} (Version: 8th June 2012) to search for spots on the full stellar disk, i.e. the parameter space for the spot position is defined to be $\theta=$ -90$^{\circ}$ to 90$^{\circ}$ with centre at 0$^{\circ}$ and $\phi=$ 0$^{\circ}$ (north pole)  to 180$^{\circ}$ (south pole) with equator at 90$^{\circ}$. However, instead of running one model to cover the full parameter range, which produced nonphysical solutions for LCs of insufficient quality and spots with low amplitudes, we narrowed the parameter space and run the model several times. In particular, we used a box of 10$^{\circ}$ times 10$^{\circ}$ which required 324 runs to cover the full visible stellar surface. To shorten the computing time we estimated the area where a spot had to be located to be occulted by the planet. Using the stellar radius in Table~2 and the planetary radius, the inclination and the semi-major axis given in Table~1 and assuming a maximum spot size of 30$^{\circ}$, we calculated a possible latitude range for the spots of $\phi=$ 20$^{\circ}$  to 120$^{\circ}$ which translates into only 180 runs of \begin{scriptsize}PRISM\end{scriptsize}. Each run consisted of 1500 iterations resulting in 270000 iterations for one transit LC. \\ We modelled all 13 \textit{CoRoT} as well as the four ground-based LCs with \begin{scriptsize}PRISM\end{scriptsize} regardless of whether spots could be identified by visual inspection. 
Since each model consists of 180 individual runs, the LC analysis resulted in 180 different fits per LC. The goodness of each fit is given as $\chi^{2}$. In all except for one case we found
a number of fits with equally good $\chi^{2}$. Instead of taking the model with the lowest $\chi^{2}$ as the final solution, which would lead to extremely underestimated error bars because of the artificial narrowing of the parameter space, we computed the mean of solutions with similar low $\chi^{2}$. The number of averaged models were chosen as 1\,$\sigma$ of the $\chi^{2}$ distribution i.e. all fits with a $\chi^{2}$ within 68.27\% of the original best-fitting model $\chi^{2}$ were taken into account. The final uncertainties are given by the standard deviation of the averaged spot parameters. For LCs with low quality and spots with low amplitudes the majority of the models yielded an equally low $\chi^{2}$, which reflects in parameters with large error bars. The analysis of the OSN LC from October 28, 2014 (epoch 862) resulted only in one best-fitting model, hence the error bars in that particular case are one order of magnitude lower than for the other LCs. In two cases (\textit{CoRoT} LC at epoch -23 and OSN LC at epoch 1104) the best-fitting model showed no sign of a spot feature.\\ Tregloan-Reed et al. (2015) published an updated version of the \begin{scriptsize}PRISM\end{scriptsize} and \begin{scriptsize}GEMC\end{scriptsize} codes. The new version includes eccentricity, the ability of modelling multiple starspots, a modifiable size of the planetary radius in pixels and a replacement of the Markov Chain Monte Carlo (MCMC) component of \begin{scriptsize}GEMC\end{scriptsize} with \begin{scriptsize}DE-MC\end{scriptsize} (Braak 2006), a combination of the genetic algorithm \begin{scriptsize}DE\end{scriptsize} (Storn \& Price 1997) with MCMC. The upgrade from a simple MCMC to \begin{scriptsize}DE-MC\end{scriptsize} however caused a drastic increase of the computing time. We conducted several tests of the two \begin{scriptsize}PRISM\end{scriptsize} versions and found them in good agreement for the parameters but with more realistic error bars for the updated version. We repeated the analysis with the updated version of \begin{scriptsize}PRISM\end{scriptsize} (Version: 6th October 2015). To reduce the computer time, we used as priors the spot parameters obtained with the first \begin{scriptsize}PRISM\end{scriptsize} version. The geometrical parameters of the transiting system ($k, \frac{R_{\mathrm{A}}}{a}+\frac{R_{\mathrm{b}}}{a}, i, T_{\mathrm{c}}$) were, however, a free parameter. To avoid nonphysical results, the limb darkening (LD) coefficients were allowed to vary $\pm$0.1 around the theoretical values for the quadratic LD law given in the table by Claret (2000) for the \textit{R}-band. Likewise, we used the LD coefficients given in Sing (2010) for the  modelling of the \textit{CoRoT} observations. The size of the planetary radius was taken as 50 pixels.

\begin{landscape}
\begin{table}
\centering
\caption{Results from the spot modelling with PRISM. $i$ -- orbital inclination, $\theta$ -- longitude of the centre of the spot in degrees from -90 to 90, $\phi$ -- latitude of the centre of the spot in degrees from 0 to 180 with equator at 90, $r_{\mathrm{spot}}$ -- spot size, $C_{\mathrm{spot}}$ -- spot contrast (1.0 equals surrounding photosphere), $N_{\mathrm{m}}$ -- Number of models used for the derivation of the spot parameters ($\chi^{2}$ within 68.27\% of the best-fitting model $\chi^{2}$)}
\begin{tabular}{lcr@{\,$\pm$\,}lr@{\,$\pm$\,}lr@{\,$\pm$\,}lr@{\,$\pm$\,}lr@{\,$\pm$\,}lr@{\,$\pm$\,}lr@{\,$\pm$\,}lc}
\hline \hline
 & Epoch & \multicolumn{2}{c}{$\frac{R_{\rm{b}}}{R_{\rm{A}}}$} & \multicolumn{2}{c}{$\frac{R_{\mathrm{A}}}{a}+\frac{R_{\mathrm{b}}}{a}$} & \multicolumn{2}{c}{$i$ [$^{\circ}$]} &  \multicolumn{2}{c}{$\theta$ [$^{\circ}$]} & \multicolumn{2}{c}{$\phi$ [$^{\circ}$]} &  \multicolumn{2}{c}{$r_{\mathrm{spot}}$ [$^{\circ}$]} & \multicolumn{2}{c}{$C_{\mathrm{spot}}$} & $N_{\mathrm{m}}$\\ \hline
CoRoT & -32  & 0.1319 & 0.0023 & 0.1596 & 0.0107 & 89.92 & 1.08 &   3.7 & 53.0 & 74.2 & 32.5 & 17.8 & 4.8 & 0.81 & 0.14 & 126 \\
CoRoT & -31  & 0.1327 & 0.0028 & 0.1645 & 0.0043 & 89.23 & 0.76 &  35.4 &  9.3 & 65.7 & 17.8 & 23.8 & 6.5 & 0.82 & 0.08 & 16 \\
CoRoT & -30  & 0.1283 & 0.0057 & 0.1620 & 0.0259 & 89.59 & 3.04 & -40.5 & 18.0 & 63.4 & 11.2 & 27.2 & 4.1 & 0.84 & 0.11 & 12 \\
CoRoT & -29  & 0.1315 & 0.0032 & 0.1571 & 0.0063 & 88.86 & 0.91 &   9.5 &  6.5 & 61.5 & 20.2 & 21.1 & 5.6 & 0.78 & 0.14 & 18 \\
CoRoT & -28  & 0.1342 & 0.0035 & 0.1560 & 0.0071 & 89.50 & 1.23 &  40.0 &  9.0 & 67.3 & 15.9 & 23.0 & 5.2 & 0.81 & 0.08 & 12 \\
CoRoT & -27  & 0.1364 & 0.0035 & 0.1688 & 0.0102 & 88.13 & 1.63 &  24.3 & 32.1 & 73.5 & 18.5 & 20.4 & 5.6 & 0.78 & 0.12 & 17 \\
CoRoT & -26  & 0.1311 & 0.0027 & 0.1608 & 0.0043 & 90.00 & 0.60 &  -0.3 & 52.3 & 69.8 & 29.1 & 17.5 & 4.5 & 0.80 & 0.14 & 178 \\
CoRoT & -25  & 0.1341 & 0.0026 & 0.1611 & 0.0031 & 89.93 & 0.42 &  23.1 & 34.7 & 68.7 & 19.8 & 21.8 & 4.5 & 0.77 & 0.16 & 18 \\
CoRoT & -24  & 0.1350 & 0.0039 & 0.1630 & 0.0123 & 89.28 & 1.74 & -13.1 &  6.1 & 66.3 & 18.0 & 25.2 & 5.0 & 0.86 & 0.04 & 16 \\
CoRoT & -23* & 0.1288 & 0.0024 & 0.1629 & 0.0074 & 89.95 & 1.65 & \multicolumn{2}{c}{} & \multicolumn{2}{c}{} & \multicolumn{2}{c}{} & \multicolumn{2}{c}{} &  \\
CoRoT & -22  & 0.1319 & 0.0058 & 0.1587 & 0.0183 & 89.96 & 2.61 &   3.6 & 51.4 & 70.4 & 30.7 & 18.0 & 5.3 & 0.80 & 0.13 & 120 \\
CoRoT & -21  & 0.1370 & 0.0041 & 0.1706 & 0.0134 & 87.39 & 1.68 & -10.2 &  1.4 & 61.9 & 14.0 & 24.6 & 3.4 & 0.88 & 0.02 & 5 \\
CoRoT & -20  & 0.1295 & 0.0038 & 0.1617 & 0.0192 & 89.91 & 2.30 &   0.6 & 55.5 & 72.9 & 26.5 & 18.5 & 5.3 & 0.82 & 0.12 & 104 \\
OSN   & 714  & 0.1395 & 0.0047 & 0.1577 & 0.0042 & 89.35 & 0.91 &  33.9 &  1.3 & 65.5 & 19.9 & 15.7 & 4.4 & 0.68 & 0.13 & 3 \\
OSN   & 862  & 0.1426 & 0.0037 & 0.1620 & 0.0082 & 89.79 & 1.43 &  -1.7 &  0.1 & 93.8 & 0.3 & 29.5 & 0.4 & 0.641 & 0.002 & 1 \\
OSN   & 872  & 0.1374 & 0.0036 & 0.1667 & 0.0108 & 87.49 & 1.58 & -20.6 &  2.1 & 61.6 & 16.7 & 16.8 & 4.9 & 0.75 & 0.11 & 7 \\
OSN   & 1104*& 0.1398 & 0.0043 & 0.1665 & 0.0089 & 88.14 & 1.96 &  \multicolumn{2}{c}{} & \multicolumn{2}{c}{} & \multicolumn{2}{c}{} & \multicolumn{2}{c}{} & \\\hline
\multicolumn{2}{c}{final} & \multicolumn{2}{c}{} & \multicolumn{2}{c}{} & 89.54 & 0.24 & \multicolumn{2}{c}{} & \multicolumn{2}{c}{} & \multicolumn{3}{c}{}  \\
\multicolumn{2}{c}{final (white)}& 0.1323 & 0.0009 & 0.1615 & 0.0018 & \multicolumn{2}{c}{} & \multicolumn{2}{c}{} & \multicolumn{2}{c}{} & \multicolumn{3}{c}{}  \\
\multicolumn{2}{c}{final (\textit{R}-band)}& 0.1398 & 0.0020 & 0.1604 & 0.0033 & \multicolumn{2}{c}{} & \multicolumn{2}{c}{} & \multicolumn{2}{c}{} & \multicolumn{3}{c}{}   \\\hline\hline
\end{tabular}
\\
$^{\ast}$ No spot feature was identified by \begin{scriptsize}PRISM\end{scriptsize}
\end{table}
\end{landscape}

\noindent The results of the spot modelling with \begin{scriptsize}PRISM\end{scriptsize} for system parameters as well as for the spot parameters of each individual transit are given in Table~3. The LCs of CoRoT-18\,b with the best-fitting spot model and a representation of the stellar disc with the starspot are shown in Fig.~1 and 2. For clarity we only show $\pm$0.1\,d around the mid-transit time of the \textit{CoRoT} transits. Only the starspot image of the best-fitting model with the lowest $\chi^{2}$ is shown here, not an image of the spot obtained by averaging all solutions with similar low $\chi^{2}$. Hence, the values of the spot parameters are different in Table~3 and Fig.~1 and 2. For the same reason, the varying size and contrast of the spots between transits is not caused by a very rapid spot evolution but it reflects the uncertainties due to the quality of the light curves (see Table~3; spot size and contrast of the same spot are in agreement within the error bars.) \\ To test the significance of the detected spot features we fitted the transits with a spot-free-model and calculated the reduced $\chi^{2}$. The difference between the spot- and the spot-free-model are shown together with the LCs in Fig.~1 and 2. In all except of two cases the reduced $\chi^{2}$ of the spot model is lower than the one of the spot-free-model. The LCs with a higher reduced $\chi^{2}$ are the ones with numerous similar good fitting models and, hence, the largest error bars in Table~3. The differences in reduced $\chi^{2}$ are, however, small, and hence the spots are only marginally detected. Only for the ground-based transits observed at OSN, the spot model fits the LCs significantly better.

\subsection{System Parameters}

The usual way to determine precise system parameters is to fit a transit model to the LCs. A suitable model accounts for planetary parameters as well as for the structure of the stellar surface i.e. stellar LD. Stellar activity, however, complicates transit modelling due to the non-homogeneous brightness distribution on the stellar surface. The planet occultation of a spotty area on a star's surface leads to a structured transit with parts of decreased depth. Unocculted spots outside the transit path lead to a decrease of the average temperature and, hence, the effective stellar radius is reduced. The observed transits are deeper than predicted by a spot-free model which results in an overestimated planet-to-star radius ratio. To understand this effect, Czesla et al. (2009) showed that it is important to take the modulation of the out-of-transit LC into account. The higher the local flux level before and after a transit, the less affected the transit is by stellar spots.\\ To account for stellar activity we tested two different approaches to determine the system parameters and compared the results. Firstly, we took the system parameters directly derived from \begin{scriptsize}PRISM\end{scriptsize}, and computed the weighted average of the values for each LC. The second approach consisted of performing simultaneous transit fitting of all \textit{CoRoT} white light and the ground-based \textit{R}-band transits with the Transit Analysis Package\footnote{http://ifa.hawaii.edu/users/zgazak/IfA/\begin{scriptsize}TAP\end{scriptsize}.html} (TAP v2.1, Gazak et al. 2012). The second approach was part of the analysis of Raetz et al. (2019) and is explained in detail in that publication. To avoid loosing information for the determination of the system parameters, Raetz et al. (2019) used the original (unbinned, non-detrended) LCs in the modelling process. The parts of the LCs where spot-features were identified by \begin{scriptsize}PRISM\end{scriptsize} were, however, removed. The results of the fitting are shown in Raetz et al. (2019) their Figure~15 and given here in Table~4.

\begin{table}
\centering
\caption{System parameters resulting from the simultaneous LC fitting with TAP (obtained by Raetz et al. 2019, values from their Table~4). The values under the horizontal line were computed from $R_{\mathrm{b}}$/$R_{\mathrm{A}}$ and $a$/$R_{\mathrm{A}}$ to allow for the comparison with the values in Table~3.}
\renewcommand{\arraystretch}{1.2}
\begin{tabular}{lc}
\hline \hline
Parameter & Value \\ \hline
    Inclination [$^{\circ}$] & 89.9 $^{+1.6}_{-1.6}$\\
           $a$/$R_{\mathrm{A}}$ & 7.013 $^{+0.078}_{-0.160}$\\
          $R_{\mathrm{b}}$/$R_{\mathrm{A}}$ (\textit{CoRoT} white light) & 0.1331 $^{+0.0014}_{-0.0013}$\\
          $R_{\mathrm{b}}$/$R_{\mathrm{A}}$ (\textit{R}-band) & 0.1410 $^{+0.0020}_{-0.0019}$\\
      Linear LD (\textit{CoRoT} white light) & 0.492 $^{+0.025}_{-0.025}$ \\
        Quad LD (\textit{CoRoT} white light) & 0.199 $^{+0.026}_{-0.026}$ \\      
      Linear LD (\textit{R}-band) & 0.384 $^{+0.041}_{-0.041}$\\
        Quad LD (\textit{R}-band) & 0.292 $^{+0.047}_{-0.048}$\\\hline
        $\frac{R_{\mathrm{A}}}{a}+\frac{R_{\mathrm{b}}}{a}$ (\textit{CoRoT} white light) & 0.1616$^{+0.0018}_{-0.0037}$\\
        $\frac{R_{\mathrm{A}}}{a}+\frac{R_{\mathrm{b}}}{a}$ (\textit{R}-band) & 0.1627$^{+0.0018}_{-0.0037}$\\\hline\hline
        \end{tabular}
\end{table}

\section{Physical properties}

As in our previous studies (e.g. Raetz et al. 2015) and following the procedures of Southworth (2009) we used the system parameters to calculate stellar, planetary and geometrical parameters of the system. In Raetz et al. (2019) we refined the physical properties using the system parameters resulting from the simultaneous LC analysis with \begin{scriptsize}TAP\end{scriptsize} only (values are given in Table~1). Here, we update the physical properties using the system parameters obtained as the weighted average of the results from \begin{scriptsize}PRISM\end{scriptsize} and \begin{scriptsize}TAP\end{scriptsize} that are given in Tables~3 and 4, respectively.\\ H\'{e}brard et al. (2011) emphasised the similarities between the CoRoT-2 and CoRoT-18 systems in terms of effective stellar temperatures,  metallicities, stellar rotation periods as well as orbital periods and planetary equilibrium temperatures. They found, however, that the inferred stellar densities for CoRoT-18 and CoRoT-2 differed slightly (Winn 2010 showed that the mean stellar density $\rho_{\mathrm{A}}$ can be derived directly from the LC). Our result is more similar to the one of CoRoT-2, that is, $1.282^{+0.031}_{-0.041}\,\mathrm{\rho}_{\mathrm{\odot}}$ vs. $1.288^{+0.035}_{-0.033}\,\mathrm{\rho}_{\mathrm{\odot}}$ for CoRoT-18 and CoRoT-2, respectively, the latter value given by Gillon et al. 2010. Hence, CoRoT-18 matches the parameters of CoRoT-2 even closer than previously estimated. \\ Our final derived physical properties that are summarised in Table~5, are consistent but more precise than the results of  Raetz et al. (2019) and are in good agreement (average deviation $\sim$0.96$\sigma$) with the values of H\'{e}brard et al. (2011) and Southworth (2012). We found the largest deviations for the stellar density $\rho_{\mathrm{A}}$, the orbital inclination $i$ and the impact parameter $b$. However, for most of the values the differences are within 2\,$\sigma$, and are therefore not significant. The discrepancies most likely arise from the different treatment of the stellar activity as well as from binning the data to 512\,s cadence by Southworth (2012), which reduces slightly the amount of information.\\ In both light curve analyses with \begin{scriptsize}PRISM\end{scriptsize} and \begin{scriptsize}TAP\end{scriptsize}, we found differences in the transit depths between the \textit{CoRoT} white light and the $R$-band observations. This discrepancy is significant with more than 5$\sigma$. One explanation might be an underestimation of the `third' light induced by contaminants in the aperture around CoRoT-18. But since CoRoT-18 is assumed to be an active, young star, this also might be explained by an increase of stellar activity over time. Our ground-based data were taken $\sim$5 years after the \textit{CoRoT} observations and, hence, a change in stellar activity seems to be plausible. An example of such activity variations were reported by Iwanek et al. (2019) for a target from the OGLE Galactic bulge data. The higher amplitude of the spot signals in the ground-based data of CoRoT-18 suggests a stronger activity. To quantify this hypothesis, we roughly estimated the change of the spot covered area between the observations in the same way as done in Raetz et al. (2014). To facilitate the calculations, we assumed that all data were taken in the same wavelength (in first approximation a reasonable choice since the $R$-band is in the middle of the \textit{CoRoT} band pass). The temporal change in planet-to-star radius ratio was transferred to a change of stellar luminosity which is due to the change of the `effective' stellar radius caused by a different spot coverage. We found a decrease of $\sim$11\% in luminosity over the 5 years. Assuming that the difference between the temperature of the spots and the temperature of the surrounding photosphere was 1700\,K (Berdyugina 2005), we derived that the stellar area covered by spots had increased by 14.2\%. This corresponds to a change in brightness of ~0.13\,mag. Similar amplitudes were found for example for the activity cycles of the young solar analogue stars LQ\,Hya (Berdyugina et al. 2002), AB\,Dor and EK\,Dra (J\"{a}rvinen et al. 2005a,b). To confirm if we detected part of such a cycle, further photometric and spectroscopic follow-up observations are  needed. \\ Thanks to the continuous \textit{CoRoT} observations, it is possible to trace a spot over several stellar rotations. Measuring the spot location on consecutive transits enables us to derive the stellar rotation period $P_{\mathrm{rot}}$ and the spin-orbit alignment $\lambda$ using simple geometry. The rotation period of CoRoT-18 was already determined from the variations in the out-of-transit global LC to be 5.4\,$\pm$\,0.4\,d by H\'{e}brard et al. (2011). Since this is only 2-3 times longer than the orbital period of the planet, it is not as straight forward to derive the stellar rotation period and the spin-orbit-alignment as for systems with $P_{\mathrm{orb}}\ll P_{\mathrm{rot}}$. In the case of fast rotating stars like CoRoT-18, spot observations do not allow us to determine the rotation period independently, but they can be used to refine the value already known through another method. If we assume a single large long-lived spot and a stellar rotation period of $\sim$5.4\,d, the planet will cross over the same starspot on average on every third transit (after a full stellar rotation). In particular, we can expect to find the same spots e.g. at \textit{CoRoT} transits at epoch -32/-29/-26, -31/-28/-25, and -30/-27/-24 (note: a possible evolution of the spots can be neglected because only the location of the centre of the circular spot is used in the calculations). For each spot we calculated $P_{\mathrm{rot}}$ and $\lambda$ for all combinations of transits (e.g.  -32 $\rightarrow$ -29,  -32 $\rightarrow$ -26, and -29 $\rightarrow$ -26). The large error bars in the spot location for a single \textit{CoRoT} transit results in large uncertainties in $P_{\mathrm{rot}}$ and $\lambda$ using only two transits. Therefore we computed the error weighted average of all values derived from all transit combinations. The results, that are given in Table~5, are in agreement with previously published values. Our measurements of $P_{\mathrm{rot}}$ are 13 times more precise while for $\lambda$ they are in the same order of magnitude. This is not surprising since $P_{\mathrm{rot}}$ is proportional only to the spot's longitude (and the time between the spot feature in two transits) which is well determined by the location of the spot anomaly in the LC. On the contrary, $\lambda$ also depends on the latitude of the spot which is correlated with the spot size and contrast both having large uncertainties.\\ With \begin{scriptsize}PRISM\end{scriptsize} it is only possible to model each transit independently from the other. Another approach to constrain the spot parameters would be a simultaneous modelling of all \textit{CoRoT} transits including the out-of transit brightness variations. To confirm our results obtained with \begin{scriptsize}PRISM\end{scriptsize}, we used the publicly available tool \begin{scriptsize}SOAP-T\end{scriptsize} (Oshagh et al. 2013a and in Boisse et al. 2012; http://www.astro.up.pt/resources/soap-t/) to simulate the LC of CoRoT-18. We fixed the stellar and planetary parameters to our results given in Tables~4 and 5. The spot parameters (longitude, latitude, spot size and contrast) of the three spots obtained in the LC analysis with \begin{scriptsize}PRISM\end{scriptsize} (Table~3) were used as input parameters for \begin{scriptsize}SOAP-T\end{scriptsize}. In addition, we placed two more spots with similar sizes on latitudes different than the transit path to reproduce the out-of-transit intensity variations. The resulting LC model shown in Fig.~4 follows the shape of the \textit{CoRoT}-LC and validates the outcome of the PRISM analysis. Note, that \begin{scriptsize}SOAP-T\end{scriptsize} does not account for spot evolution and therefore only the first part of the LC can be fitted reasonably well. A main result of \begin{scriptsize}SOAP-T\end{scriptsize} is that at least several spots of a similar size are required to reproduce the full LC (transit and out-of-transit) of CoRoT-18. Moreover, the LC simulation points to additional spots outside of the transit path and, hence, yields useful information to understand how spotted the surface of this star is. This finding is in agreement with similar studies on young stars with transiting planets. Bruno et al. (2016) showed, for example, that six to nine activity features of which one or two are found to cross the transit chord, are required to reproduce the full LC of CoRoT-2. Because of the high level of degeneracy and correlation of the parameters in the simulation, we can not put tighter constraints on the spot parameters. \\ With our refined stellar rotation period we estimated the stellar age via gyrochronology by using the calibrated gyrochronology relation given in  Angus et al. (2015). The result of $194^{+265}_{-85}$\,Myrs also supports the young star hypothesis.

\begin{figure}
  \centering
  \includegraphics[width=0.5\textwidth,angle=270]{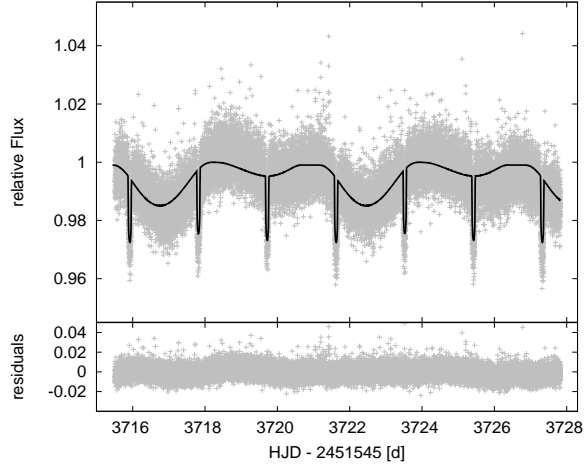}
  \caption{Part of the normalised \textit{CoRoT} white light LC. The model obtained with \begin{scriptsize}SOAP-T\end{scriptsize} with the three spots in the transit path obtained by \begin{scriptsize}PRISM\end{scriptsize} and two additional dark spots outside of the transit path is given by the black solid line. The bottom plot shows the residuals of the fit.}
\end{figure}

\begin{table}
\centering
\caption{Physical properties of the CoRoT-18 system derived from LC modelling. The physical properties are calculated using the weighted average of the system parameters determined with PRISM and TAP. The values of Raetz et al. (2019) determined by LC fitting with TAP only, are given for comparison in Table~1.}
\begin{tabular}{lr@{\,$\pm$\,}l|lr@{\,$\pm$\,}l|lr@{\,$\pm$\,}l}
\hline
\multicolumn{3}{c|}{planetary parameters} & \multicolumn{3}{c|}{stellar parameters} & \multicolumn{3}{c}{geometrical parameters}\\ 
Parameter & \multicolumn{2}{c|}{This work} &  Parameter & \multicolumn{2}{c|}{This work} &  Parameter & \multicolumn{2}{c}{This work} \\ \hline 
$R_{\mathrm{b}}$  [R$_{\mathrm{Jup}}$] & 1.141 & $^{0.034}_{0.035}$ & $R_{\mathrm{A}}$  [R$_{\mathrm{\odot}}$] & 0.883 & $^{0.024}_{0.025}$ & $a$  [au] & \multicolumn{2}{c}{0.0288}\\
$M_{\mathrm{b}}$  [M$_{\mathrm{Jup}}$] & 3.30 & $^{0.19}_{0.19}$ & $M_{\mathrm{A}}$  [M$_{\mathrm{\odot}}$] & 0.88 & 0.07 & &  \multicolumn{2}{c}{$\pm$0.0008} \\
$\rho_{\mathrm{b}}$  [$\mathrm{\rho}_{\mathrm{Jup}}$] & 2.08 & $^{0.22}_{0.23}$ & $\rho_{\mathrm{A}}$  [$\mathrm{\rho}_{\mathrm{\odot}}$] & 1.28 & $^{0.03}_{0.04}$ & $i$  [$^{\circ}$] & 89.55 & $^{0.24}_{0.24}$ \\
log\,$g_{\mathrm{b}}$ & 3.801 & $^{0.016}_{0.017}$ & log\,$g_{\mathrm{A}}$ & 4.490 & $^{0.013}_{0.015}$ & $b$ & 0.055 & $^{0.029}_{0.029}$   \\
$T_{\mathrm{eq}}$ [K] & 1487 & $^{19}_{19}$ & $P_{\mathrm{rot}}$ [d] & 5.19 & 0.03 & $\lambda$ [$^{\circ}$] & 6 & 13 \\ 
$\Theta$ & 0.189 & $^{0.019}_{0.019}$ & & \multicolumn{2}{c|}{} & & \multicolumn{2}{c}{} \\ \hline
\end{tabular}
\end{table}

\section{Transit Timing}

When dealing with active stars it is very important to pay special attention to the determination of the transit times. Spot features in a transit LC can lead to transit timing variations and, hence, to a false-positive detection of additional non-transiting bodies in the system (e.g. Alonso et al. 2009; Sanchis-Ojeda et al. 2011). Oshagh et al. (2013b) found that, depending on the size and the location of the spot, transit timing variations with amplitudes up to 200\,s can be produced when a typical Sun-like spot is present.\\ Raetz et al. (2019) determined the mid-transit times (spot features were removed before the analysis) and refined the ephemeris. To check the robustness of the transit times determination we repeated the analysis using the LCs with the spot features. For a valid comparison we also used the original (unbinned, non-detrended) LCs as done by Raetz et al. (2019). We found a maximum difference between the transit times of $\sim$60\,s which is well within the individual error bars. \\ Furthermore we estimated the order of magnitude of the timing variations caused by occulted spots following the calculations in Sanchis-Ojeda et al. (2011). For a spot with an amplitude $A_{\mathrm{S}}$, a duration $T_{\mathrm{S}}$ and a midpoint $t_{\mathrm{S}}$ the induced timing variations can be approximated by
\begin{equation}
\Delta t_{\mathrm{Spot}}\approx \frac{\frac{1}{2}A_{\mathrm{S}}T_{\mathrm{S}}(t_{\mathrm{S}}-T_{\mathrm{c}})}{\left(R_{\mathrm{b}}/R_{\mathrm{A}}\right)^{2}T } 
\end{equation}
where $T$ is the time between the ingress and egress and $T_{\mathrm{c}}$ is the mid-transit time. Using the spot locations derived by \begin{scriptsize}PRISM\end{scriptsize}, we estimated the expected timing variation. In most cases the influence of the spots on the transit times is well within their individual error bars. Only for one LC (OSN LC at epoch 714) with a high amplitude spot feature that is located far away from mid-transit, the estimated timing variation exceeds the individual errors. The final results for the transit times  (converted to BJD$_{\mathrm{TDB}}$ with the online tool\footnote{http://astroutils.astronomy.ohio-state.edu/time/utc2bjd.html} of Eastman et al. 2010), the transit time difference between spotted and spotless LCs as well as the estimated timing variation are given in Table~6.

\begin{table}
\centering
\caption{Transit times for all transits of CoRoT-18\,b. $T_{\mathrm{c,Spot}}$ -- mid-transit time of LCs with spot features, $T_{\mathrm{c}}$ -- mid-transit time of the spot removed LCs ( for the values see Raetz et al. 2019, their Table~13), $\Delta t_{\mathrm{Spot}}$ -- estimate of the timing variations caused by occulted spots.}
\renewcommand{\arraystretch}{1.1}
\begin{tabular}{ccr@{\,$\pm$\,}lr@{\,$\pm$\,}lc}
\hline \hline
 Telescope & Epoch & \multicolumn{2}{c}{$T_{\mathrm{c,Spot}}$ [BJD$_{\mathrm{TDB}}$]} & \multicolumn{2}{c}{$T_{\mathrm{c}}-T_{\mathrm{c,Spot}}$  [s]} & $\Delta t_{\mathrm{Spot}}$ [s] \\ \hline \hline
CoRoT & -32  &  2455260.922563 &  $^{0.00076}_{0.00081}$ &  -3 & $^{ 92}_{ 96}$   &  -6 \\
CoRoT & -31  &  2455262.823533 &  $^{0.00088}_{0.00088}$ &  -4 & $^{108}_{106}$   &   6 \\
CoRoT & -30  &  2455264.721683 &  $^{0.00140}_{0.00140}$ &   9 & $^{165}_{165}$   &  -2 \\
CoRoT & -29  &  2455266.623083 &  $^{0.00140}_{0.00150}$ & -26 & $^{154}_{171}$   &   3 \\
CoRoT & -28  &  2455268.522373 &  $^{0.00085}_{0.00082}$ &  -2 & $^{100}_{ 97}$   &   7 \\
CoRoT & -27  &  2455270.423983 &  $^{0.00082}_{0.00081}$ &   1 & $^{ 97}_{ 98}$   &   7 \\
CoRoT & -26  &  2455272.324033 &  $^{0.00072}_{0.00074}$ &  -1 & $^{ 86}_{ 89}$   &   6 \\
CoRoT & -25  &  2455274.223372 &  $^{0.00094}_{0.00092}$ & -15 & $^{116}_{112}$   &  11 \\
CoRoT & -24  &  2455276.123682 &  $^{0.00110}_{0.00110}$ &  43 & $^{134}_{128}$   & -14 \\
CoRoT & -23* &  2455278.023582 &  $^{0.00100}_{0.00097}$ &  \multicolumn{2}{c}{}  &     \\
CoRoT & -22  &  2455279.923942 &  $^{0.00086}_{0.00089}$ &  10 & $^{104}_{105}$   &  -6 \\
CoRoT & -21  &  2455281.822282 &  $^{0.00130}_{0.00140}$ &  60 & $^{165}_{165}$   &  -6 \\
CoRoT & -20  &  2455283.723132 &  $^{0.00082}_{0.00082}$ & -15 & $^{ 98}_{ 98}$   &  -3 \\
OSN   & 714  &  2456678.389660 &  $^{0.00059}_{0.00060}$ &  12 & $^{ 64}_{ 65}$   &  62 \\
OSN   & 862  &  2456959.602891 &  $^{0.00079}_{0.00081}$ &  10 & $^{ 85}_{ 87}$   &   3 \\
OSN   & 872  &  2456978.605017 &  $^{0.00110}_{0.00110}$ &  26 & $^{128}_{128}$   & -46 \\ 
OSN   & 1104*&  2457419.424813 &  $^{0.00095}_{0.00100}$ &  \multicolumn{2}{c}{}  &     \\
\hline \hline
\end{tabular}
\\
$^{\ast}$ No spot feature was identified by \begin{scriptsize}PRISM\end{scriptsize}\\
\end{table}

\section{Discussion and Conclusions}

CoRoT-18\,b is one out of three planets discovered by the \textit{CoRoT} mission that were found to orbit a young star. It is a massive Hot Jupiter that orbits its host star with a period of less than 2\,d on an eccentric orbit. Planets around young stars have been discovered mainly with the direct imaging technique and, hence, their orbital separations are large. There are only of the order of ten (massive) Hot Jupiters in a tight orbit around a young star known so far. Although CoRoT-18\,b is one of these rare objects, it had not been yet a target for photometric follow-up monitoring or dedicated observing campaigns. \\ We collected four high-precision ground-based LCs and re-analysed 13 available \textit{CoRoT} transits. Most of the LCs show starspot features during the transit that we analysed in more detail. In most LCs the spot features are only marginally detected (S/N between 0.3 and 2.5). The ground-based transits show, however, significant stellar spot in the transit path. This is not surprising as the average spot amplitudes for the \textit{CoRoT}-transits are lower than for the ground-based LCs. The outcome of our starspot model is shown in Fig.~1 and 2. From the modelling of the spots we could derive the stellar rotation period and the spin-orbit-(mis)alignment. We found that the stellar spin axis of CoRoT-18 is aligned with the orbital axis of its planet which confirms the measurement of the Rossiter-McLaughlin effect reported by  H\'{e}brard et al. (2011). Our value for the stellar rotation period is 13 times more precise than the value derived from the variations of the global LC. In summary, we could confirm the previously published value of the spin-orbit-(mis)alignment by using an independent method, and by using a prior on the stellar rotation period we could refine the period value derived through another method.\\ To reduce systematic errors caused by spot anomalies, transit measurements at times when the flux level of the overall LC is the highest should be used to determine accurate system parameters. However, none of the observed \textit{CoRoT} transits happened at a maximum of the LC and all transits close to a maximum showed spot features. Consequently, the determined system parameters as well as the derived physical properties might be systematically biased. Therefore, to minimise any biases, we have compared two different methods and calculated the weighted average of the results. We found that the physical properties derived in our study slightly deviate from the values of H\'{e}brard et al. (2011) and Southworth (2012), most likely due to the different treatment of the stellar activity as well as a different approach to data binning. We can confirm the results of Raetz et al. (2019) and found that the systematics that could be introduced by the spots did not affect them, and that they are robust against different modelling methods. \\ Its highly variable LC, the fast rotation period and the high lithium abundance, are strong indicators that that CoRoT-18 is a young star. However, the age determination is highly uncertain as other age indicators point to very different ages. Stellar evolution models (e.g. PARSEC isochrones,  Bressan et al. 2012) yield either a very young or an old age while through gyrochronology (using the relation given in Angus et al. 2015) an age of $194^{+265}_{-85}$\,Myrs is obtained. Lanza (2015) showed that tides can spin up planet host stars. Therefore, the age estimated by gyrochronology could be underestimated in the presence of a massive Hot Jupiter. Furthermore, gyrochronology is not a reliable method to estimate the age of stars younger than about 1 Gyr as Sun-like stars in young open cluster exhibit a bimodal distribution of their rotation period (e.g., Barnes 2003; Meibom et al. 2009, 2011). However, the measured rotation period of CoRoT-18 is consistent with an age less than 1\,Gyr according to a recent model of stellar rotation evolution (see Fig. 3 in Gondoin 2017) that reproduces the bimodal distribution of stellar rotation observed in young open clusters. CoRoT-18 is also probably older than 30\,Myrs as young pre-main sequence stars have $R'_{\mathrm{HK}}$ indices close to $10^{-4}$ ( e.g. Gondoin et al. 2012). Although we cannot exclude an old or a very young age for CoRoT-18 (which seems to be unlikely as pointed out by Guillot \& Havel 2016), our observation and the analysis of the Ca II H and K lines in the HARPS spectra support the young star hypothesis with an age of several hundred Myrs.\\ By comparing the transit depths of the \textit{CoRoT} white light and $R$-band observations we found differences that are significant with 5$\sigma$. Our estimates showed that this discrepancy can be explained by a change of stellar activity over time as part of a possible activity cycle. Similar amplitudes for activity cycles were found for several young solar analogue stars, and, hence, our estimate is another indication that CoRoT-18 is a young star. The confirmation of the activity cycle hypothesis requires further follow-up observations.\\ When translating the $\sim$5-day rotation period into an Rossby number, we found that the measured emission flux in the core of the Ca II lines of CoRoT-18 follows the empirical relation between the $R'_{\mathrm{HK}}$ index and the Rossby number established by Mamajek \& Hillenbrand (2008). The determined Rossby number places CoRoT-18 in the non-saturation regime in the rotation-activity relation for late-type stars (relation between Rossby number and X-ray to bolometric luminosity ratio, Wright et al. 2011). The resulting high value of the X-ray surface flux leads to a stellar wind with a mass loss rate up to 600 times higher than the solar wind. \\ With an orbit semi-major axis of 0.0288 au, the planet CoRoT-18\,b is not only subjected to the huge photospheric flux and gravity field of its nearby mother star, but also to an intense flux of high energy photons and particles including flares or coronal mass ejections. Such an environment has most likely a drastic impact on its outer atmosphere (e.g. Owen \& Jackson 2012).

\Acknow{SR acknowledge support from the People Programme (Marie Curie Actions) of the European Union's Seventh Framework Programme (FP7/2007-2013) under REA grant agreement No. [609305]. MF acknowledges financial support from grants AYA2014-54348-C3-1-R, AYA2011-30147-C03-01 and AYA2016-79425-C3-3-P of the Spanish Ministry of Economy and Competivity (MINECO), co-funded with EU FEDER funds. The present study was made possible thanks to observations obtained with \textit{CoRoT}, a space project operated by the French Space Agency, CNES, with participation of the Science Program of ESA, ESTEC/RSSD, Austria, Belgium, Brazil, Germany and Spain. }

\end{document}